\documentclass[twocolumn,showpacs,floats,superscriptaddress,10pt]{revtex4}
\usepackage{graphicx}
\usepackage{amsmath}

\begin{document}

\title{Valley-dependent Brewster angles and Goos-H\"{a}nchen effect in
strained graphene}
\author{Zhenhua Wu}
\affiliation{SKLSM, Institute of Semiconductors, Chinese Academy of Sciences, P.O. Box
912, 100083, Beijing, China}
\author{F. Zhai}
\affiliation{Department of Physics, Zhejiang Normal University, Jinhua 321004, China}
\author{F. M. Peeters}
\affiliation{Department of Physics, University of Antwerp, Groenenborgerlaan 171, B-2020
Antwerpen, Belgium}
\author{H. Q. Xu}
\affiliation{Division of Solid State Physics, Lund University, Box 118, S-22100 Lund,
Sweden}
\author{Kai Chang}
\email{kchang@semi.ac.cn}
\affiliation{SKLSM, Institute of Semiconductors, Chinese Academy of Sciences, P.O. Box
912, 100083, Beijing, China}

\begin{abstract}
We demonstrate theoretically how local strains in graphene can be tailored
to generate a valley polarized current. By suitable engineering of local
strain profiles, we find that electrons in opposite valleys ($K$ or $%
K^{\prime }$) show different Brewster-like angles and Goos-H\"{a}nchen
shifts, exhibiting a close analogy with light propagating behavior. In a
strain-induced waveguide, electrons in $K$ and $K^{\prime }$ valleys have
different group velocities, which can be used to construct a valley filter
in graphene without the need for any external fields.
\end{abstract}

\maketitle

\emph{Introduction.} In recent years, graphene, a single layer of carbon
atoms arranged in a hexagonal lattice has shown abundant new physics and
potential applications in carbon-based nanoelectronic devices.~\cite%
{Novoselov2,Zhang,Neto} The novel properties arise from the linear energy
dispersion and the chiral nature of electrons at the $K$ and $K^{^{\prime }}$
valleys of the Brillouin zone.~\cite{Neto} Mathematically, the Dirac
equation describing the motion of massless quasiparticles is very similar to
the Helmholtz equation for an electromagnetic wave, except for the negative
energy spectrum. This remarkable fact could make it possible to observe
optical-like electron propagating behavior and to construct Dirac
electron-optical devices with graphene.~\cite{Cheianov} The mean-free path
of electrons in graphene can approach microns at room temperature, making
electrons behave ballistically in a graphene microstructure.~\cite%
{Ozyilmaz,Morozov} In such a ballistic regime, the scattering of electrons
by potential barriers can be understood by comparing with the reflection,
refraction and transmission of electromagnetic waves in inhomogeneous media.
Therefore, graphene is a testbed to examine optical-like phenomena of Dirac
fermions. Several studies have been devoted to such optical-like electron
behaviors using external electric bias.~\cite%
{Cheianov,Park,Darancet,Goos,Chen,Beenakker,Ghosh,Zhao,Concha,Williams}
These proposals require to deposit a metallic gate above graphene, which is
an additional complication.

\begin{figure}[b]
\centering \includegraphics [width=0.9\columnwidth]{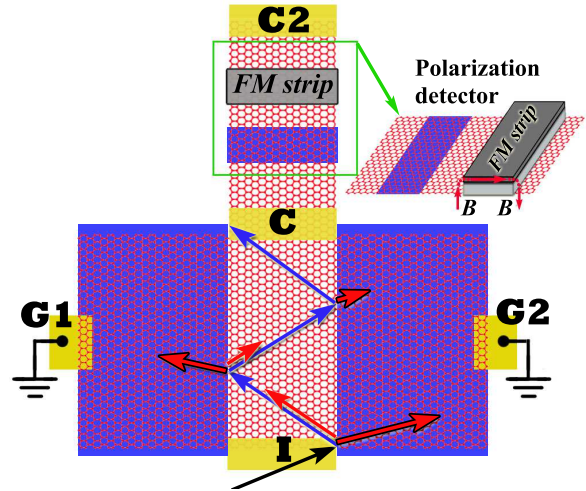}
\caption{Schematic of a strained four-terminal guiding device including
contacts \textbf{I} (injector), \textbf{C} (collector), and \textbf{G1},
\textbf{G2} (electrical ground 1, 2). Strain exists in the shaded regions.
The black arrow denotes the unpolarized incident electron beam. The red
(blue) arrows stand for the $K$-valley ($K^{\prime }$valley) polarized
electron beams. A valley detector as proposed in Ref.~\protect\cite{Zhai}
could be connected to contact \textbf{C} to measure the polarization. The
detector is also a graphene-based device with a substrate strain and a
ferromagnetic (FM) strip.}
\label{fig:model}
\end{figure}

The strain effect in graphene provides a new way to manipulate electron
transport without external fields. The two valleys in graphene are related
by time reversal symmetry and act in much the same way as electron spin in
spintronics. The valley degree of freedom in graphene might be used to
control the characteristics of graphene-based devices, referred to as
valleytronics.~\cite{Xiao,Pereira,Rycerz} Realization of such valleytronics
requires an effective and robust scheme to produce valley polarized
currents, although several schemes about valley filters were proposed
utilizing, e.g., the edge profile of graphene nanoribbons,
the trigonal warping effect in the energy spectrum, and intense
irradiation on a bilayer graphene.~\cite{Rycerz,ZZZhang,JMPJ,Abergel} In this letter, we propose a
simple and effective way to realize some optical-like behaviors of electrons
and valley filters by using the strain effect alone. We find that electrons
in opposite valleys can be perfectly transmitted or totally reflected in the presence of strain.  A quantum waveguide formed between two strained regions (see Fig.~\ref{fig:model}) can confine electrons in it and the reflected beam is shifted
laterally along the interface by a distance with respect to the incident
beam, as the Goos-H\"{a}nchen (GH) effect~\cite{Goos} in optics. We
demonstrate that the valley-dependent GH effect in graphene results in
different group velocities for electrons in $K$ and $K^{\prime }$ valleys.

\begin{figure}[tbp]
\centering \includegraphics[width=\columnwidth]{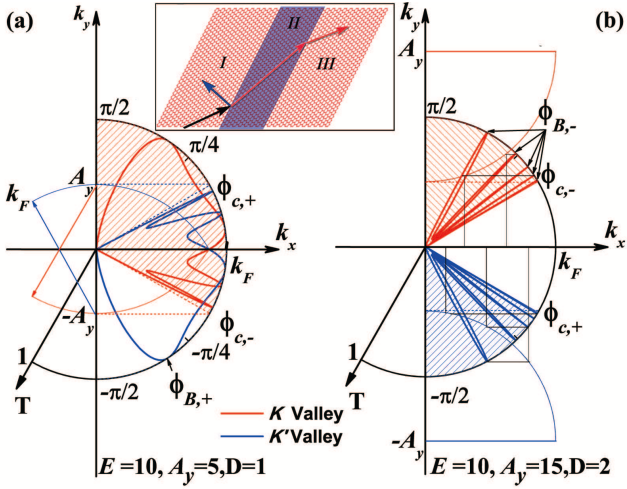}
\caption{The transmission probability $T$ as a function of the incident
angle $\protect\phi $. In the strained region, the wave vectors of electrons
in $K$ ($K^{\prime }$) valley satisfy $k_{x,\protect\xi }^{2}+(k_{y}+\protect%
\xi A_{y})^{2}=E^{2}$ as indicated by the red (blue) arc curve. The red
(blue) dashed region indicates the transmission window for electrons in the $%
K$ ($K^{\prime }$) valley. (a) $E=10$, $A_{y}=5$, $D=1$. (b) $E=10$, $%
A_{y}=15$, $D=2$.}
\label{fig:2}
\end{figure}

\emph{Model.} Mechanical strain in graphene can be described by a gauge
vector potential. To illustrate the operating principle of the proposed
device, we take a simplified gauge field which is equivalent to a
delta-function-like antiparallel magnetic double barrier perpendicular to
the graphene monolayer, but with opposite direction at $K$ and $K^{\prime }$
valleys.~\cite{Bao,Guinea} The Landau gauge $\boldsymbol{A}=(0,A_{y},0)$ is
adopted in our calculation. The length scale of the spatial variation of the
pseudo gauge vector potential is much larger than the lattice spacing of
graphene, which implies that intervalley scattering is weak at low-energy regions. The low-energy electrons can be well described by the
effective Hamiltonian $H=v_{F}\boldsymbol{\sigma }^{(\prime )}\cdot (%
\boldsymbol{p+}\xi \boldsymbol{A}^{i}\boldsymbol{/}v_{F})+V^{i}$,~\cite%
{Pereira} where the superscript $i$ indicates the different
regions (see Fig.~\ref{fig:2}), $v_{F}$ is the Fermi velocity, $%
\boldsymbol{\sigma }^{(\prime )}$ $(\boldsymbol{\sigma }^{\prime }=-%
\boldsymbol{\sigma }\ )$ are the pseudospin Pauli matrices, $\boldsymbol{p}$
is the electron momentum, $V^{i}$ is the electrostatic potential in the
region $i$, and $\xi \boldsymbol{=}\mp 1$ labels $K$ and $K^{\prime }$
valleys. In the calculation, we introduce the dimensionless length and
energy units: $l_{B}=(\hbar /eB_{0})^{1/2}$, $E_{0}=$ $\hbar v_{F}/l_{B}$
(which are $81.1$ nm and $7.0$ meV for a typical pseudomagnetic field $%
B_{0}=0.1$ T). 

\emph{Brewster-like Angle.} We start by investigating electron transmission
through a region of uniform uniaxial strain with width $D$ as shown in the
inset of Fig.~\ref{fig:2}. Here, we set $V^{i}=0$ for simplicity since the
key physical mechanism to introduce the valley-dependent Brewster angle is
the gauge vector $\xi \boldsymbol{A}$. The translational invariance along
the \textit{y} direction gives rise to conservation of transverse wave
vector $k_{y}$, and thus the solutions can be written as $\psi (x,y)=\psi
(x)e^{ik_{y}y}$. In the strained region, the longitudinal wave vector $%
k_{x,\xi }=\sqrt{(E^{2}-(k_{y}+\xi A_{y})^{2}}$ has a different dependence
on the vector potential for the two valleys denoted by $\xi =\mp 1$, since
the strain-induced\ pseudomagnetic fields have different signs for the two
valleys. The reflected amplitude reads%
\begin{equation}
{\small r_{\xi }=\frac{\text{sin}(k_{x,\xi }D)(\text{sin}\theta _{\xi }-%
\text{sin}\phi )(\text{sin}\phi -i\text{cos}\phi )}{\text{sin}(k_{x,\xi
}D)(1-\text{sin}\theta _{\xi }\text{sin}\phi )+i\text{cos}(k_{x,\xi }D)\text{%
cos}\theta _{\xi }\text{cos}\phi },}  \label{coeff}
\end{equation}%
where the incident and refractive angles are defined as $\phi \equiv arc\sin
(k_{y}/E)$ and $\theta _{\xi }\equiv arc\sin [(k_{y}+\xi A_{y})/E]$. These
expressions are valid for incident angle $\phi <\phi _{c,\xi }\equiv \arcsin
(|k_{y}+\xi A_{y}|/E)$. $\phi _{c,\xi }$ is the critical angle for total
reflection in optics and $\sin \theta _{\xi }/\sin \phi =(k_{y}+\xi
A_{y})/k_{y}\equiv n$ gives Snell's law for transmitted electrons. Note that
when $k_{y}(k_{y}+\xi A_{y})<0$, the refractive index $n$ of the strained
graphene is negative just like for a metamaterial with a negative refractive
index. Importantly, one can tune the refractive index $n$ mechanically in
quite a large range, which is not so for the metamaterials. When $\phi >\phi
_{c,\xi }$, electron beams can be totally reflected at such a strained
barrier, since the longitudinal wave vector $k_{x,\xi }$ becomes imaginary
accounting for the occurrence of evanescent modes in the barrier.

Figure~\ref{fig:2} shows the angular dependence of the transmission
probability $T_{\xi }=1-|{\small r_{\xi }}|^{2}$, which exhibits a
remarkable valley-dependence. It is obvious that $T_{\xi }(\phi )=T_{\bar{\xi%
}}(-\phi )$, where $\bar{\xi}=-\xi $, i.e., the transmission of
electrons in $K$ and $K^{\prime }$ valleys shows a mirror symmetry
about $\phi =0$. This feature is guaranteed by the time reversal
symmetry in strained graphene. In the presence of uniform uniaxial
strains, the wave vectors of electrons in $K$ valley satisfy
the relationship $k_{x,-1}^{2}+(k_{y}-A_{y})^{2}=E^{2} $ as\
indicated by the red arc curve. When $|k_{y}-A_{y}|>E$,
(equivalent to $\phi \in \lbrack -\pi /2,\phi _{c,-}]$), the wave
vector $k_{x}^{^{\prime }}$ in the strained region becomes
imaginary, indicating the appearance of evanescent modes, thus
transmission is totally blocked. The critical angle for total
reflection is given by Snell's law, $\phi _{c,-}=\arcsin
((A_{y}-E)/E)$. Thus the transmission window of electrons in
$K$ valley is restricted to the red shadowed region ($\phi \in
\lbrack \phi _{c,-},\pi /2]$), as shown in Fig.~\ref{fig:2}(a). In
contrast, the transmission of electrons in $K^{\prime }$
valley shows mirror symmetry behavior with respect to that in
$K$ valley. Similarly, the critical angle for electrons in
$K^{\prime }$ valley is given by $\phi _{c,+}=\arcsin
((E-A_{y})/E)$. As a consequence, there exist mirror symmetric
transmission windows for $K$ and $K^{\prime }$ electrons that can
be completely separated by increasing the strain gauge fields
[Fig.~\ref{fig:2}(b)]. One can see in Fig.~\ref{fig:2} that for
some specific incident angles $\phi _{B}$, resonant peaks that
reach perfect transmission exist in different windows for
electrons in $K$ or $K^{\prime }$ valley. This can be
readily checked by analyzing the zeros of the reflected amplitude $r$
corresponding to $\sin (k_{x,\xi }D)=0$. The perfect tunneling
peaks are fully coincident for $k_{x,\xi }=n\pi /D$ [see
Fig.~\ref{fig:2}(b)]. The
number of resonant peaks increases with the incident energy $E$ and the width $%
D$ of the strained region. Note that if such resonant peaks for one valley $%
\xi $ are located in the transmission gap for the other valley $\bar{\xi}$,
we can obtain valley-$\xi $ polarized transmitted electron beams and valley-$%
\bar{\xi}$ polarized reflected beams. These characteristic angles are
analogous to the so-called Brewster angle in optics, at which \emph{s}- and
\emph{p}-polarized light are produced. We can propose a valley polarization
device that is straightforward and analogous to the Glan-Thompson optical
polarizers. The strained stripe leads to perfect transmission of one valley
component and total reflection of the other valley components when electron
beams are incident at the Brewster-like angles $\phi _{B}$.

\emph{Valley-dependent Goos-H\"{a}nchen Effect.} In analogy with
total reflection of light at the interface between two materials
with different refractive indexes, the electronic GH effect
describes the shift $\sigma _{GH}$ of the reflected electron beam
for total internal reflection at an interface, along the
transverse direction. The GH shift could be positive or negative
which is determined by external electric~\cite{Beenakker} and/or
inhomogeneous magnetic fields~\cite{Chen}, an analogue with the
optical GH effect in normal materials or metamaterials. A uniform
uniaxial strain in graphene can lead to a valley-dependent GH
shift $\sigma _{GH,\xi }=(2k_{y}+\xi A_{y})/(k_{x}\kappa _{\xi
})$, where $\kappa $ is the modulus of the imaginary wave vector
in the strained region in the case of total internal reflection.
For a waveguide formed in between two strained stripes as shown in
Fig.~\ref{fig:model}, the transmission gap at each interface can
lead to the confinement of electrons in the channel associated
with multiple total internal reflections. The difference between
the tunneling forbidden regions for electrons in different valleys
gives rise to valley-dependent electron propagation along the
channel. In the channel, the electronic GH shifts at each
interface are accumulated during multiple reflection processes.

\begin{figure}[tbp]
\centering
\begin{minipage}[c]{0.49\columnwidth}
\centering
\includegraphics[width=\columnwidth]{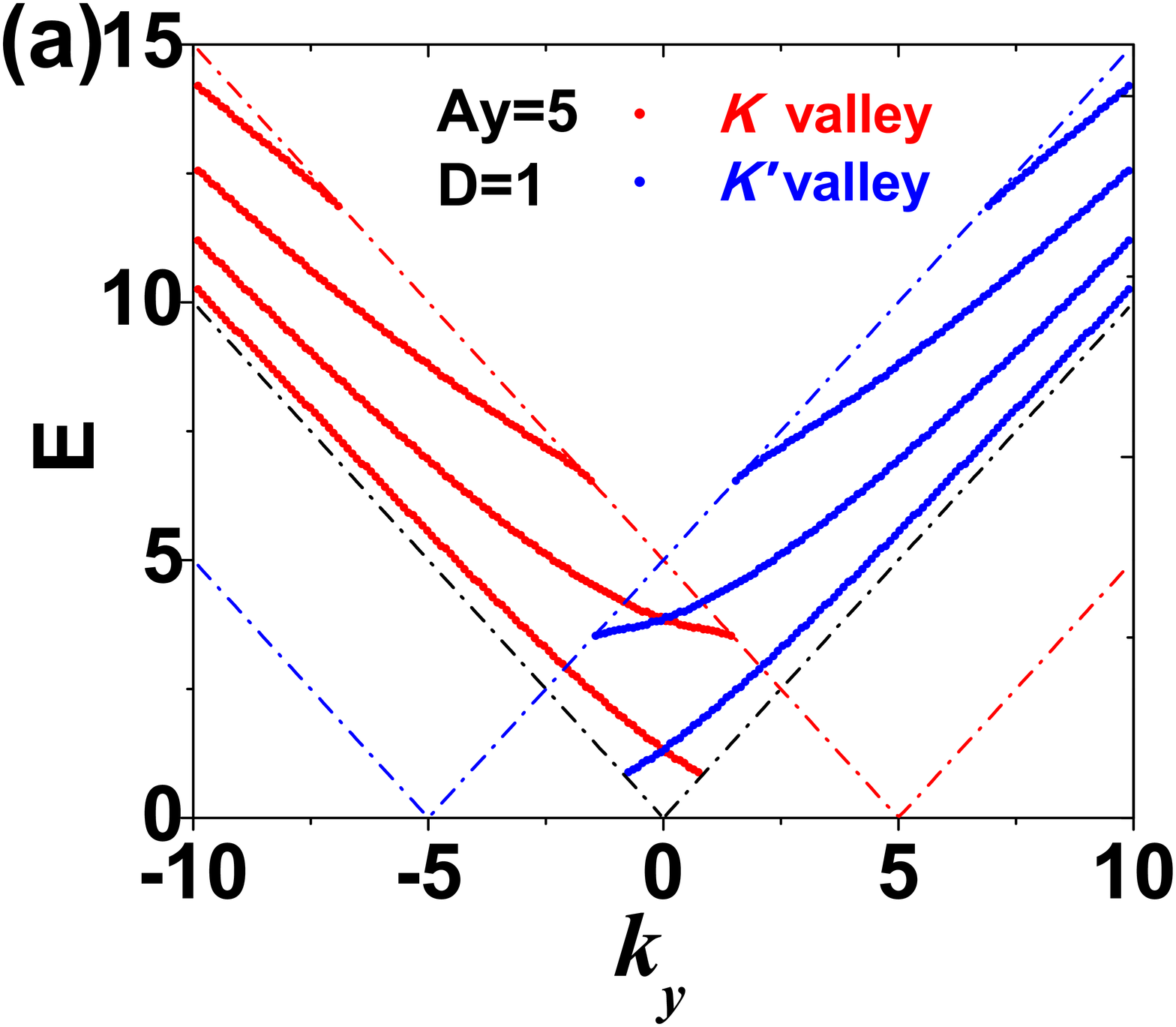}
\end{minipage}
\begin{minipage}[c]{0.49\columnwidth}
\centering
\includegraphics[width=\columnwidth]{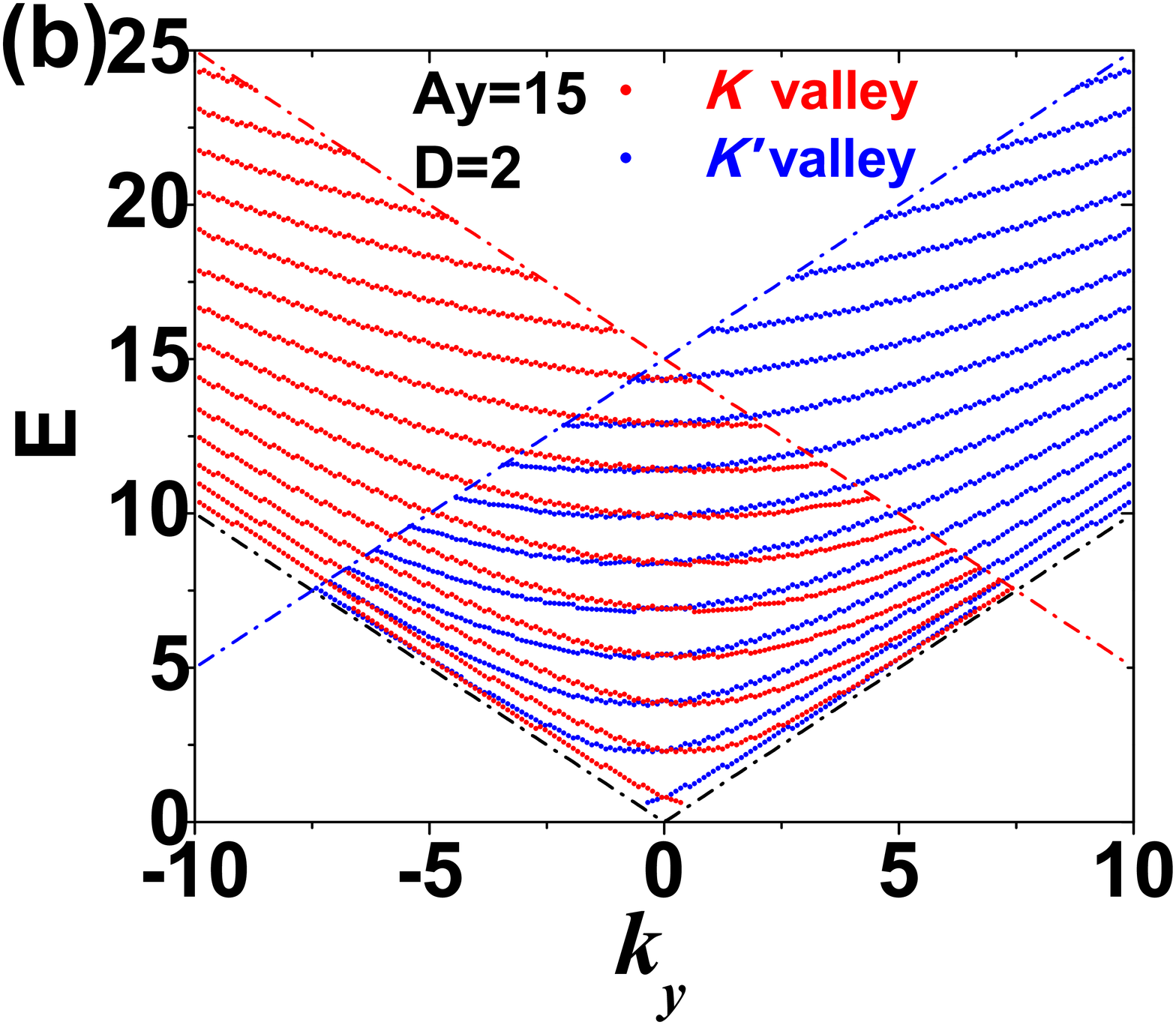}
\end{minipage}
\caption{The energy spectra for the lowest channel modes in the
stain-induced waveguides for (a) $A_{y}=5$, $D=1$, and (b) $A_{y}=15$, $D=2$%
. }
\label{fig:3}
\end{figure}

The GH effect is found when there is total internal reflection, and thus
there are bound states localized in the channel between the two interfaces
as shown in Figs.~\ref{fig:3}(a) and \ref{fig:3}(b). The bound states exist
only in the region between the curves given by the equations $|k_{y}+\xi
A_{y}|=|E|$ and $|k_{y}|=|E|$, i.e., the transmission gap as we discussed
above. The spectrum of each valley ($K$ or $K^{\prime }$) alone is
asymmetric, but for both valleys ($K$ and $K^{\prime }$), the total spectra
are mirror symmetric with respect to the transverse wave vector $k_{y}$, as
required by time reversal symmetry. That is very different from bound states
in a pure electric waveguide in graphene, where the bound states are
valley-independent and always mirror symmetric with respect to $k_{y}=0$.~%
\cite{Beenakker} The number of bound states increases with the
strain-induced gauge filed $A_{y}$. Three typical regions in the energy
spectrum are found: i) $E<A_{y}/2$, i.e., $\phi _{c,\mp }=\pm \pi /2$, the
incident electrons from both valleys are always totally reflected at the
interfaces. The bound states for $K$- or $K^{\prime }$-valley electrons
coexist for all incident angles. The electrons in different valleys have
different group velocities determined by the slope of the energy dispersion
relation. Thus such a waveguide may be used to separate the electrons in
different valleys after passing a sufficient long channel. Notice that there
are several local minima in the dispersion relation indicating a vanishing
group velocity. Thus electrons in $K$ or $K^{\prime }$ valley will be
trapped in the waveguide, which can be used to construct a valley memory
device. ii) $A_{y}/2<E<A_{y}$, i.e., $\pi /2>\phi _{c,-}>\phi _{c,+}>-\pi /2$%
, the incident electrons from both valleys are confined in the channel for
incident angles $\phi \in \lbrack \phi _{c,+},\phi _{c,-}]$. The
transmission in such an angular window is similar to that in situation i).
For $\phi <\phi _{c,+}$ (or $\phi >\phi _{c,-}$), only the bound states for $%
K$ valley (or $K^{\prime }$ valley) electrons appear, while electrons in $%
K^{\prime }$ valley (or $K$ valley) are able to penetrate into the strained
region and eventually disappear from the channel region. So we can produce a
valley-polarized current in such a strain-induced waveguide. iii) $A_{y}<E$,
i.e., $\pi /2>\phi _{c,+}>\phi _{c,-}>-\pi /2$. In this case, there is no
coexistence of bound states for $K$ or $K^{\prime }$ valley. The
transmission properties are similar as that in case ii) for $\phi <\phi
_{c,+}$ and $\phi >\phi _{c,-}$. Note that this valley-dependent transport
property can not be realized in normal electric/magnetic waveguides proposed in previous works (see
Refs.~\cite{Chen,Beenakker}).

\begin{figure}[t]
\centering \includegraphics [width=\columnwidth]{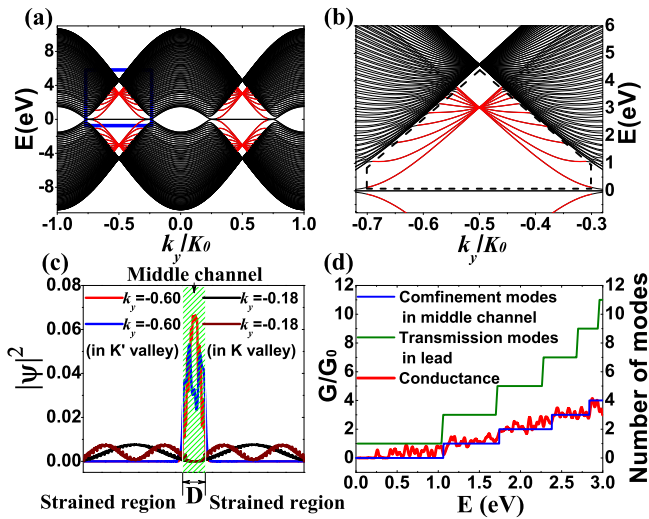}
\caption{(a) Energy dispersion of a strained graphene ribbon with
width N=140 (in unit of honeycombs).~\cite{ZZZhang} Lateral
strains are applied on the ribbon except the middle part as shown in Fig.~\protect\ref{fig:model}.
\textbf{$K_{0}$} is a reciprocal lattice vector. (b) A blowup of
the box region in panel (a). (c) Density
distributions of electrons in the lowest two forward-going modes with $%
k_{y}=-0.60K_{0}$ in $K^{\prime }$ valley and $k_{y}=-0.18K_{0}$ in $K$
valley. (d) Conductance versus incident energy. The blue (green) curve gives
the number of propagating modes in the channel (lead).}
\label{fig:4}
\end{figure}

Finally, we focus on how the valley-polarized transmission features shown
above are reflected in the conductance $G$. Recently, Williams et al.~\cite{Williams} reported fiber-optic guiding of electrons in graphene. Differential
resistance can be measured using a standard lock-in technique and thus guiding efficiency can be extracted.
Our proposal depicted in Fig.~\ref{fig:model} is a valley-related version of their setup. Here the guiding modes serve as valley-filtered modes while the bulk modes are unpolarized. To distinguish these valley-filtered modes from bulk modes, a
higher guiding efficiency is desired. High guiding efficiency can be achieved
by applying a uniform magnetic field as reported in Ref.~\cite{Williams}.
But in this letter we propose a different scheme, where an applied mechanical strain
behaves like a delta-function-like pseudomagnetic field, with opposite
direction at the $K$ and $K'$ valleys. We performed
numerical simulations of electrical conduction in a tight-binding model~\cite%
{ZZZhang} of a strained graphene stripe. The dispersion relation $E(k_{y})$
is plotted in Figs.~\ref{fig:4}(a) and \ref{fig:4}(b). The guiding modes are
formed in the interval of the bulk states as shown in the dashed pentagon in
Fig.~\ref{fig:4}(b). Note that when $E<3$ eV, all the forward-going guiding
modes are located in $K^{\prime }$ valley, while the forwarding-going modes
in $K$ valley are solely bulk modes. The density distributions of electrons
in the lowest two forward-going modes in $K$ and $K^{\prime }$ valleys are
shown in Fig.~\ref{fig:4}(c). Electron guiding modes (in $K^{\prime }$
valley) and bulk modes (in $K$ valley) are separated spatially in the
channel and the strained region, respectively. The conductance shows good
correspondence with the number of guiding modes rather than the much larger
number of transmission modes in the leads [see Fig.~\ref{fig:4}(d)]. This is
because electrons in bulk modes in the lead leak out of the channel and
eventually disappear together with those in bulk modes in the strained
region via the electric grounds \textbf{G1} and \textbf{G2}. The reason for
the oscillation of the conductance is that the stripe has a finite width and
some bulk modes in the strained region still affect the conductance. One
would expect that the conductance will approach the guiding mode steps very
well when the strained region are electrical grounded. The guiding modes can
effectively carry current from the injector \textbf{I} to the collector
\textbf{C} and thus dominate the channel conductance. As we
demonstrated that the guiding modes are valley-filtered,
the desired valley polarized current can be measured in collector \textbf{C}%
. A valley detector as proposed in Ref.~\cite{Zhai} can be connected to the
contact \textbf{C} to identify the polarization. The detector shows a large
valley-resistance ratio, in analogy to the giant magnetoresistance effect.
Thus valley polarized current could be identified by measuring the
conductance at terminal \textbf{C2}.

In summary, We demonstrated theoretically how a strain-induced gauge field
can be tailored to generate valley-polarized transport in a single layer of
graphene. Our results show that the propagating behavior of electrons in
graphene exhibits deep analogies with light as a consequence of the massless
linear Dirac dispersion and the chirality of electron states. Most
importantly, the gauge fields induced by strain can lead to valley-dependent
transport phenomena, e.g., the Brewster angles and the Goos-H\"{a}nchen
effect. Electrons in a waveguide structure formed by two strained stripes
propagate along the channel with different velocities at different valleys.
This feature shed new light on constructing graphene-based valleytronic
device by using only strain, without the need for any external fields.

This work was supported by the NSF of China, the Swedish International Development Cooperation Agency (SIDA), and the Belgian Science Policy (IAP).

\end{document}